# Sharding and HTTP/2 Connection Reuse Revisited: Why Are There Still Redundant Connections?

Constantin Sander, Leo Blöcher, Klaus Wehrle, Jan Rüth

Communication and Distributed Systems, RWTH Aachen University, Aachen, Germany

{sander,bloecher,wehrle,rueth}@comsys.rwth-aachen.de

## ABSTRACT

HTTP/2 and HTTP/3 avoid concurrent connections but instead multiplex requests over a single connection. Besides enabling new features, this reduces overhead and enables fair bandwidth sharing. Redundant connections should hence be a story of the past with HTTP/2. However, they still exist, potentially hindering innovation and performance. Thus, we measure their spread and analyze their causes in this paper. We find that 36 % - 72 % of the 6.24 M HTTP Archive and 78 % of the Alexa Top 100k websites cause Chromium-based webbrowsers to open superfluous connections. We mainly attribute these to domain sharding, despite HTTP/2 efforts to revert it, and DNS load balancing, but also the Fetch Standard.

## 1 INTRODUCTION

Internet standards of the past decade, such as HTTP/2 [2] and HTTP/3 [3], have paved the Web's way to use a single transport connection. While HTTP/1.1 needs multiple concurrent connections to achieve parallelism, its successors can multiplex content over a single connection. A single connection has many advantages on paper, e.g., connection-establishment overheads such as the 3-way-handshake + additional TLS handshakes or growing the congestion window (slow start) diminish. Further, it decreases resource use as fewer connections have to be maintained, especially at content-providers with many users. But focussing on a single connection also offered further innovation potential within HTTP: header compression and resource prioritization became viable, and new features such as server push could be tested. Research found that these new features can significantly improve web performance when used correctly [25, 27, 28]. Here, multiple connections can even degrade effectiveness as, for instance, prioritization does not span across connections and priorities lose their meaning.

However, simply updating a webserver to use the most recent standard does not revert previous performance tricks. E.g., domain sharding, i.e., spreading content over various subdomains, was seen as a valid way to gain more connections and parallelism in the age of HTTP/1.1. The structures created by these practices are still present [25], as even CDNs do not reduce redundancy if domain sharding is structurally still enforced by spreading resources over domains. A recognized problem, the HTTP/2 and HTTP/3 standards[1] try to revert domain sharding with their Connection Reuse mechanism when a subdomain resolves to an IP for which a connection is already established. Nonetheless, a 2016 study [13] showed that browsers still tended to open multiple HTTP/2 connections to the same domain. Roughly one-third of all flows were duplicates although HTTP/2 should use a single connection [2, Section 9.1] – the root causes remained vague.

In this work, we revisit the occurrence of redundant connections w.r.t. HTTP/2 to discover if the problem persists and, more importantly, identify why Connection Reuse is ineffective and what can be done for rectification. We do not solely focus on domain sharding but generally look at connections that could have been avoided. To this end, our study is two-fold: We analyze data provided in the HTTP Archive [16] (visiting millions of websites per month to collect web statistics) to get an idea of the scale of redundant connections and further perform additional measurements on the Alexa Top 100k to pinpoint and attribute redundant connections. Specifically, our paper contributes and finds the following:

- We elaborate why Connection Reuse, a seemingly simple reversal of domain sharding, can miss its target in practice.
- We present and apply a method to quantify the causes of redundant connections of Chromium browsers.
- Redundancy still persists; 2.26 M to 4.49 M of the 6.24 M HTTP Archive and 77.88 k of Alexa Top 100k websites are affected.
- DNS load-balancing is the leading cause, then privacy-driven Web standards and domain sharding with separate certificates.
- We find that only few parties cause the majority of redundancy.

## 2 MULTIPLE CONNECTIONS AND HTTP

Modern web pages consist of a multitude of resources [11]. However, HTTP/1.1 [8] uses a single TCP connection to send one resource after another. I.e., a delayed resource, e.g., due to database accesses, can delay subsequent resources hindering the rendering process.

---

---

[1]It is easier to include a workaround in a standard that is eventually implemented than to get operators to change their practices.

## 2.1 HTTP/1 – Parallel Connections

Hence, browsers open six or more parallel TCP connections [10, 13, 27] to achieve more parallelism with HTTP/1.1. Website operators employed domain sharding [10] to stretch these limits by spreading resources over more domains causing additional connections, e.g., images can be moved to a subdomain (img.example.tld). However, each connection has its costs. For instance, all connections have to be maintained on client- and server-side. While negligible for clients, the overhead for servers maintaining several thousand connections can easily build up. Similarly, latency penalties occur, e.g., with TCP, 1 RTT is spent on connection establishment, increasing to 2 or 3 RTTs when TLS is added. Additionally, congestion control (CC) slow starts with every new connection, which adds several RTTs of latency until the full throughput can be achieved.

## 2.2 HTTP/2 and HTTP/3 – One Connection

As latency is essential for web performance, HTTP/2 [2] and its successor HTTP/3 [3] follow the goal to avoid new connections. They use multiple streams that are multiplexed over a single connection to allow for parallel transfers of resources. For this, HTTP/2 implements stream semantics on top of TCP, while HTTP/3 uses the novel transport QUIC and its integrated streams. This enables better efficiency on server-side, but also innovations such as header compression and fine-grained scheduling of data – to prioritize important resources over less critical ones – on protocol-level.

*2.2.1 Effects of Redundancy.* These features were developed assuming that the Web would use only a single connection. When data remains split across connections, overheads are not saved, and prioritization and header compression cannot boost performance.

In that regard, Bocchi et al. [4] find that fewer connections usually increase the QoE with HTTP/2 in real-world settings. Wang et al. [26] find generally worse page load times (PLTs) with multiple connections for HTTP/2's predecessor SPDY. However, they find improved PLTs for high packet loss. Similarly, Goel et al. [9] find that multiple connections can worsen HTTP/2's PLTs for many small objects, but improve them for few, large objects – especially when loss is high – which they ascribe to growing the cumulative congestion window (ccwnd) faster. While not discussed by the authors, this, however, allocates more than a fair bandwidth share compared to a single connection. Manzoor et al. [14] also find multiple connections to be beneficial for HTTP/2 under high loss. They additionally attribute this to better exploitation of ECMP load-balancing and TCP's head-of-line (HOL) blocking during packet loss, pausing all HTTP/2 streams. However, the authors also find that Google QUIC (predecessor of IETF QUIC and HTTP/3) alleviates the HOL issues and surpasses HTTP/2 in many scenarios. Marx et al. [15] also find HTTP/2 to benefit from multiple connections and higher ccwnds for large resources. Nevertheless, they see that HTTP/2's PLTs for many small resources worsen when using multiple connections and that header compression is less effective as the compression dictionary has to be bootstrapped again.

We argue that with QUIC and HTTP/3, allowing easily tunable CC and removing suffering from HOL blocking, a single connection might be the desired state in all scenarios to best exploit its features and performance. Moreover, fewer connections mean fewer competing CCs and potentially better fairness. Also, content-providers probably benefit from fewer connections increasing the efficiency of their servers due to reduced connection maintenance overhead.

*2.2.2 Connection Reuse.* Nevertheless, techniques such as domain sharding intend to open multiple connections in all cases, and Varvello et al. [25] find that websites switching to HTTP/2 do not adapt but still place resources across domains.

HTTP/2 [2] hence specifies Connection Reuse to revert domain sharding: Requests for domain D may be sent over an existing connection A if D resolves to the same destination IP that A is using (+ matching ports) and if A's TLS certificate includes D (e.g., via Subject Alternative Name, SAN). I.e., an image residing on the img.subdomain-shard hosted on the same server as the root document can reuse an existing connection. HTTP/3 inherits this mechanism.

However, it is unknown whether this mechanism is effective. Varvello et al. [25] find fewer connections when switching to HTTP/2 but also that domain sharding is still used. They do not analyze whether these domains were correctly reused or whether more connections could have been avoided. On the other hand, Manzoor et al. [13] find multiple HTTP/2 connections for the same domain (in total 33%), i.e., connection reuse seems to be ineffective here. Later, the authors note that a fixed Chromium bug introduced this behavior [14]. However, we still see redundant connections for which the precise reasons and the extent are unknown.

We thus recognize a need to again look at Connection Reuse, domain sharding and why redundant connections still exist in the real world. Hence, we devise a methodology to analyze redundant connections but we will first present different combinations of website and network structure as well as browser behavior that we have identified to lead to redundant connections.

## 3 CAUSES OF MULTIPLE CONNECTIONS

Connection Reuse [2] depends on two factors: the destination IP and the domain. If an already opened connection uses the same destination IP as a new request, the connection may be reused if, additionally, its certificate includes the domain. Thus, a browser opening connections for a given domain and IP can have different reasons, which we visualize in Figure 1 and present in the following. **Unknown 3rd Party Requests:** If the IP differs and open connections do not include the new domain in their certificates, as for third party resources, where the third party has not been contacted before, a new connection has to be opened. We argue these cannot be avoided in the HTTP context, but would require a redesign of a website such that we ignore them in the following.

**Different Certificates (CERT):** Also, if the IP is the same, the domain might not be included in previous certificates. I.e., if operators use domain sharding and different certificates for their domains, HTTP/2 still opens new connections.

**Different IPs (IP):** Vice versa, the domain can be included in previous certificates, but the request's destination IP differs. I.e., really distributed resources, but also domain sharding with one certificate and differing IPs in the DNS can still open new connections.

**Fetch Standard (CRED):** Even if both factors match, browsers can refuse Connection Reuse when following the WHATWG Fetch Standard [24]. Depending, e.g., on a request's tainting type (changes, e.g., for cross-origin resource sharing / CORS requests such as font downloads across domains) and credentials mode, the Fetch Standard decides whether credentials (such as cookies) should be included in the request. A browser then only reuses a connection if its previous requests also included (or vice versa did not include) credentials (cf. §4.6, §4.7, §2.5 of [24]). Otherwise, the existing connection would be tainted with identifying information, or vice versa, a new request would be tied to previously used credentials [22]. This privacy-enhancing measure can, of course, lead to a new connection to the same IP and SAN-included domain, as is discussed by the author [22]. While Chromium implements this mechanism [12], its necessity is discussed [22] and, e.g., Firefox does not follow it [23]. In essence, the discussion in [22] revolves around the privacy effect of opening a new connection to the same server as identifying information could also be injected into Ajax request URLs and servers could also map user identities via, e.g., user IPs such that the actual privacy improvements are little to non-existing.

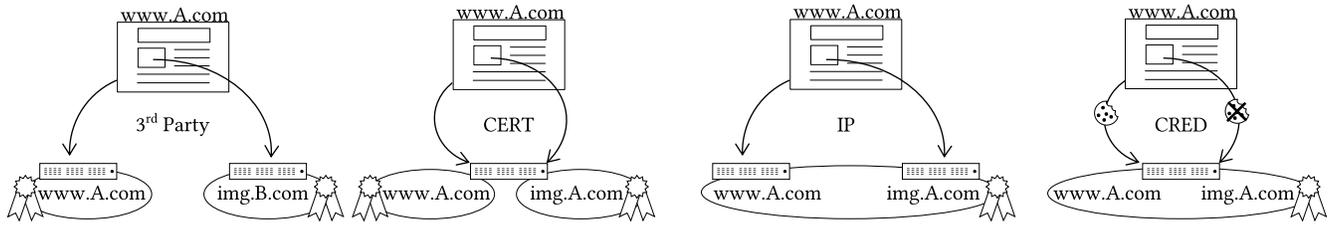

**Figure 1: Visualization of the four root causes that lead browsers to create a new HTTP/2 connection.**

**Exception: Explicitly Excluded Domains.** Additionally, Web servers can announce no support for a domain via HTTP Status 421 [2] or HTTP ORIGIN Frames [18], disabling connection reuse.

In the following, we present our methodology that allows attributing redundancies to these root causes.

## 4 METHODOLOGY

To analyze real-world websites for redundant connections w.r.t. HTTP/2, we rely on Chromium browsers to record their connections when visiting these websites. We then analyze and classify the connections accordingly.

### 4.1 Connection Analysis

For the analysis, we group connections w.r.t. HTTP/2 (in the following also HTTP/2 sessions) by their IP to find causes CERT and CRED, and, for IP and CRED, group their initially used domain name and certificate SANs of previous connections. Domains which web servers explicitly exclude, e.g., via HTTP status 421, are ignored. Moreover, we intercept the corner case of same-initial-domain requests on different IPs. Otherwise, these would be classified as IP, but only happen when CRED forbids reuse and multiple IPs are announced via DNS. We hence mark these cases as CRED.

Inherently, connections can be redundant due to multiple causes. For example, when we see four successively opened same-IP connections, where #1 and #3 use certificate A and #2 and #4 use B, we find three redundant connections in total but attribute them at time of connection establishment to three times type CERT (#2 is redundant to #1, #3 is redundant to #2, #4 is redundant to #1 and #3) and two times type CRED (#3 is redundant to #1, #4 to #2).

### 4.2 Chromium-based Connection Data

To gather the actual session information, we base our analysis on Chromium / Chrome browsers visiting websites. We focus on Chrome as it makes up around 2/3 of the browser market share [21]. Additionally, more and more browsers build on Chromium. In total, we use two different sources: We rely on the HTTP Archive's [16] desktop browser crawls from April 2021 (6.24 M websites) and also use Chromium to visit the Alexa Top 100k of April 20th, 2021.

*4.2.1 HTTP Archive.* The HTTP Archive [16] crawls the top websites of the Chrome User Experience Reports using Chrome, providing aggregated statistics and HAR files with detailed page-load information. For every website, the landing page is loaded 3 times and the HAR file for the median load time is saved. We parse these HAR files to identify HTTP/2 requests on the same sessions (by socket / connection ID) to reconstruct the HTTP/2 session lifecycle. We ignore HTTP/3 / QUIC requests as these all have socket ID 0, i.e., we cannot distinguish between the connections. Moreover, HAR files only give request-level information, i.e., we can determine the start time of a connection by the first request but cannot determine the end time precisely. Hence, we evaluate two cases: One (endless), where connections are kept open, and one (immediate), where connections are closed after the last request. The latter is probably atypical, as long flows are desired, but we still evaluate it to give a lower bound, while the former likely overestimates flow durations and redundancy counts.

*4.2.2 Own Measurements.* To gather more detailed information, we complement the HTTP Archive data with additional measurements. We leverage Browsertime [20] to visit the first 100k domains of the Alexa Top 1M with Chromium 87.0.4280.88. As URLs, we use the domains in the Alexa list (second level domains and deeper) preceded with https:// and load every landing page once. We set a page-load timeout of 300s, do not ignore certificate errors, and disable QUIC to focus on HTTP/2 and avoid switching between HTTP/3 and HTTP/2 after observing an alt-svc header. To achieve reproducibility, we disable Chromium field trials, that would otherwise randomly enable experimental features or parameterizations. We then collect Chromium's NetLog [19] files giving more details on low-level connection events (e.g., start and end) and stitch these events together to gather a precise view of the session lifecycle for analyzing it as described before.

**Ethics.** Our Chromium measurements are conducted from within our university's network on a dedicated IP inside our measurement subnet. Aiming to minimize impact, we set up a reverse DNS entry for our IP, hinting at our research context, and provide a website explaining our measurements and how to opt out. Abuse e-mails are handled correspondingly.

### 4.3 Limitations

As is typical for measurements, our approach is limited: E.g., providers blocking our measurements (we provide instructions on our website) can skew results for our vantage point.

Further, we are limited by Chromium's features: We filter HTTP status 421 [2] (see Sec. 3) in our measurements to not wrongly classify these cases as unwanted redundancy but cannot consider ORIGIN Frames [18] as these are not implemented in Chromium [17]. I.e., if Web servers signal to reuse a connection for other domains via this frame, Chromium and in turn our analysis do not react.

Also, we only review landing pages, which can show different behavior than internal pages [1]. However, the HTTP Archive focuses solely on landing pages and we aim at a broader overview of many different websites.

Additionally, cookie accept-banners are not clicked, such that further requests/connections are potentially missed due to missing consent. Furthermore, caching effects are ignored as the caches of the browsers are reset after each visit. Also, as described before, we ignore HTTP/3 requests in the HTTP Archive, as these requests lack information to attribute them to individual connections; and in our measurements to avoid potentially switching between HTTP/2 and HTTP/3 in between. We would expect that the results are comparable, as the IP and CRED case would occur equally and certificates are probably also shared for use with HTTP/2 and HTTP/3.

Aside from that, our classification can misclassify redundant connections: E.g., with newly opened connections, browsers can switch IPs if multiple IPs are announced. Hence, cause CRED can be misclassified as IP and CERT as a third party. To specifically distinguish between CRED and the remaining cases, we conduct one more measurement described later (cf. Section 5.3.3).

Lastly, we are limited by logging inconsistencies in the HTTP Archive's HAR files: We ignored 26.93 k requests with 0 as socket ID, 1.30 k / 653 requests with missing / inconsistent IPs, 66.75 M / 273.49 k / 124.37 k requests with invalid HTTP request methods / versions / statuses, and 14 requests with an incorrect page reference. 2.22 M requests did not provide SSL certificates, which we use for SAN extraction, and 11.12 M / 172.73 M requests were HTTP/3 or HTTP/1 requests. Further, we filtered 9 HAR files with invalid certificates and one without request IDs. In total, 69.12 M of 401.63 M HTTP/2 requests and 5.33 M websites were affected by these inconsistencies, but all the aforementioned information are used by our analysis to distinguish between HTTP/2 requests and missing or inconsistent entries can skew our results such that we conservatively exclude these. I.e., our HTTP Archive results potentially underestimate requests and open connections.

### 5 RESULTS

In the following, we present our results of the analysis of redundant HTTP/2 connections. In total, we analyzed 6 242 688 websites of the HTTP Archive (from April 2021), of which 5 883 212 open at least one HTTP/2 connection. Moreover, we measured the Alexa Top 100k websites from April 20[th] twice in May 2021. The first time, we followed the Fetch Standard, while the second time, we ignore its connection pool credentials flag. We found 18 282 / 18 309 sites to be unreachable for the first / second measurement run. We review the intersection of websites for comparability, consisting of 81 553 sites that all opened at least one HTTP/2 session. For ease of readability, we round percentages to integer numbers.

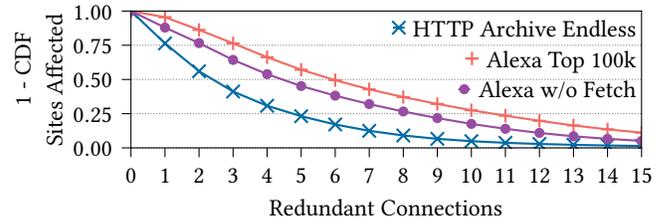

**Figure 2: Distribution of connections per website**

### 5.1 Websites with Redundant Connections

Depending on the assumed connection duration, we find between 4 493 097 (endless) to 2 263 751 (immediate) of the 5 883 212 HTTP/2 sites (76 % / 38 %) in the HTTP Archive to open at least one redundant connection (also shown in Table 1). From the Alexa Top 100k measurements, we find 77 878 HTTP/2 sites (95 %) opening redundant connections. If we had assumed endless connection duration (here we know the exact durations), these numbers do not change a lot, and we get 77 898 sites with redundant connections. As such, we find that connections are rather long-lived with a median lifetime of 122.2s for those connections that close prior to our test ending (3.5%). Given the small difference here, we continue by regarding the endless connection case for the HTTP Archive data.

Figure 2 shows the distribution of sites in relation to their redundant connections. For the HTTP Archive (×), around 50 % of all sites open at least two redundant connections. For the Alexa Top 100k (+), around 50 % open at least six.

We explain the differences, i.e., the higher number of affected websites and connections for our own measurements, by (1) the different sets of websites, (2) our measurements including also requests which we could not analyze in the HTTP Archive (cf. Section 4.3, e.g., H3 requests making up 2 % of all requests), and (3) our measurements being conducted from a different vantage point seeing different load-balancing (cf. Section 5.3.1 + Appendix A.4).
**Takeaway.** *Redundant connections are no story of the past. 76 % of all HTTP Archive and 95 % of all Alexa Top 100k HTTP/2 websites open redundant connections. The majority of websites open even more than one additional connection, potentially hindering HTTP/2 features and degrading performance.*

### 5.2 Causes of Redundant Connections

To further understand why Connection Reuse is ineffective and how it can be supported, we next analyze failure reasons in Table 1. Please note again, the sum of causes can exceed the number of sites and connections (cf. Sec. 4.1).

We can see in both datasets that differing IPs are the major cause of missing Connection Reuse. 70 % / 88 % of websites of the HTTP Archive / Alexa Top 100k with at least one HTTP/2 request are affected. Connection-wise, 22 % / 28 % of connections are affected. We later analyze which parties are causing these redundancies.

| Cause | IP | Cert | HAR Endless | | HAR Immediate | | Alexa Endless | | Alexa | | Alexa w/o Fetch | |
|---|---|---|---|---|---|---|---|---|---|---|---|---|
| | | | Sites | Conns. | Sites | Conns. | Sites | Conns. | Sites | Conns. | Sites | Conns. |
| CERT | = | ≠ | 592.95 k | 885.40 k | 299.71 k | 390.56 k | 14.17 k | 23.98 k | 14.13 k | 23.63 k | 13.88 k | 19.30 k |
| IP | ≠ | = | 4.10 M | 13.85 M | 1.73 M | 4.59 M | 71.87 k | 460.35 k | 71.86 k | 458.46 k | 71.35 k | 416.91 k |
| CRED | = | = | 2.54 M | 3.91 M | 1.35 M | 1.65 M | 64.98 k | 138.22 k | 64.83 k | 132.67 k | 0 | 0 |
| | Redund. | | 4.49 M | 17.33 M | 2.26 M | 6.42 M | 77.90 k | 579.61 k | 77.88 k | 574.85 k | 71.70 k | 429.44 k |
| | Total | | 5.88 M | 63.55 M | 5.88 M | 63.55 M | 81.55 k | 1.65 M | 81.55 k | 1.65 M | 81.55 k | 1.50 M |

**Table 1: Counts of occurring causes of redundant connections and affected websites**

In contrast, CRED, so Connection Reuse actually working but browsers following the Fetch standard still opening new connections, could be easily reduced by few, well-known browser vendors given that its privacy benefits are dubious [22] and some vendors already refrain its implementation. It is hence fortunate that it affects the second-most websites (43 % / 79 %). However, it affects far fewer connections than IP, namely only 6 % / 8 %.

Lastly, the CERT cause, i.e., domain sharding with disjunct certificates, constitutes a minority with 10 % / 17 % affected websites and 1 % / 1 % affected connections. Similar to IP, it is unclear whether few or many parties cause this effect.

**Takeaway.** *We can see that most sharded connections are of cause IP, so domain sharding with divergent IPs, for which it is unclear who causes / can rectify it. In contrast, roughly half of all websites are affected by cause CRED, pronouncing the effect of the Fetch Standard, which can be easily adapted by few browser vendors. Non-overlapping certificates affect a smaller minority of websites and connections.*

## 5.3 Unraveling Causes for Redundancy

To dive deeper into who is causing redundancy, we next analyze the causes in detail, beginning with the majority.

### 5.3.1 IP: SAN included domains with differing IPs.
Table 2 shows the top 4 origins for redundant connections of cause IP with their potentially reusable connections' origins.

We can see that the results for the HTTP Archive and our measurements overlap well, but we can, of course, see differences in their rank of occurrence (↑) and previous connections. We attribute this to (1) the different websites but also to (2) the differing vantage points and measurement times and hence differing load which influences DNS-based load-balancing (We further look into the influence of DNS-based load-balancing in Appendix A.4).

Mainly, two parties are involved: Google, with Google Analytics being the top origin in both datasets and domains indicating advertisements, and Facebook. These two parties also occur as top ASs being involved in IP redundant connections (cf. Appendix A.2)

We checked website samples where the first entry occurred and found the following behavior: The website downloads a Javascript from googletagmanager.com (GT), which then downloads a script from google-analytics.com (GA), loading further resources. Both domains were included in their respective connection's certificate but resolved to slightly different IPs in the same /24 network in our tests. When requesting the GA script on the GT IP, we received the same resource, i.e., only a single connection should suffice.

We attribute this effect to unsynchronized DNS load-balancing of both domains (cf. Appendix A.4). I.e., one domain (seen from our vantage point) is independently load-balanced from another. While we use our own recursive resolver, load-balanced resolvers with differing caches can also cause this effect.

| Origin | HTTP Archive | | Alexa 100k | |
|---|---|---|---|---|
| | ↑ | Conns. | ↑ | Conns. |
| www.google-analytics.com | 1 | 2.25 M | 1 | 52.31 k |
|   prev: www.googletagmanager.com | | 2.12 M | | 36.93 k |
| www.facebook.com | 2 | 1.52 M | 4 | 25.05 k |
|   prev: connect.facebook.net | | 1.51 M | | 25.02 k |
| googleads.g.doubleclick.net | 3 | 615.45 k | 6 | 17.03 k |
|   prev: pagead2.googlesyndication.com | | 398.41 k | | 10.19 k |
| pagead2.googlesyndication.com | 4 | 606.59 k | 7 | 16.99 k |
|   prev: googleads.g.doubleclick.net | | 418.69 k | | |
| www.google.de | 12255 | 6 | 2 | 27.74 k |
|   prev: www.gstatic.com | | | | 18.61 k |
| apis.google.com | 82 | 14.25 k | 3 | 26.10 k |
|   prev: www.gstatic.com | | | | 25.91 k |

**Table 2: Top 4 origins, their redundant connections, rank (↑) and reusable previous connections for cause IP.**

We observed a similar result for samples of the Facebook case, where connect.facebook.net (CFB) and www.facebook.com (WFB) resolve to slightly different IPs in the same /24 network. A Javascript is loaded from CFB which initiates loading a 1x1px GIF from WFB. The script from CFB can also be requested on WFB's IP, however not vice-versa. I.e., there seems to be a real resource distribution in the background in that direction. Nevertheless, ignoring potential scalability issues, resolving CFB to WFB would reduce redundancy.

The same can be found for hotjar.com (Web Analytics) which was the next non-Google nor Facebook case (cf. Appendix Table 12). An exception are the Wordpress statistics tools and extensions of wp.com (c0.wp.com, stats.wp.com) which point to different IPs in different /24 networks which are not interchangeable.

We cannot rule out setup mistakes or that content is truly distributed for the other origins, but load-balancing can similarly explain those results. Adjusting the domains to point to the same CNAME to exploit recursive resolver caches to route requests to the same connection could help here or usage of Anycast CDNs [5] (which could point each customer to exactly the same IP for every of their domains). Alternatively, adoption of the HTTP Origin Frame [18] could be a sleek way to reroute requests to the same connection and avoid redundancy.

**Takeaway.** *Major drivers for IP cases are embedded tracking / ad scripts from Google and Facebook. I.e., only a few parties have to*

| Certificate Issuer | HTTP Archive | | | Alexa 100k | | |
|---|---|---|---|---|---|---|
| | ↑ | Conns. | Domains | ↑ | Conns. | Domains |
| Let's Encrypt | 1 | 302.47 k | 63.13 k | 2 | 6.43 k | 2.36 k |
| Google Trust Services | 2 | 282.63 k | 3.24 k | 1 | 8.75 k | 239 |
| DigiCert Inc | 3 | 130.07 k | 14.86 k | 3 | 4.04 k | 651 |
| Sectigo Limited | 4 | 38.21 k | 16.78 k | 6 | 782 | 345 |
| Cloudflare, Inc. | 5 | 29.70 k | 11.55 k | 7 | 760 | 358 |
| GlobalSign nv-sa | 6 | 22.72 k | 2.28 k | 4 | 1.07 k | 296 |
| Amazon | 8 | 16.22 k | 2.31 k | 5 | 841 | 347 |

**Table 3: Top 5 certificate issuers w.r.t. redundant connections of cause CERT and unique domains.**

*adapt here to reduce the total amount of redundancy significantly. Exemplarily, we found this effect to be rooted in unsynchronized DNS-based load-balancing.*

*5.3.2 CERT: SAN excluded domains with equal IPs.* Continuing, we shed light on CERT cases to see why redundant connections occur due to disjunct certificates on the same hosts. We begin by analyzing who issued the certificates; Table 3 shows the top certificate issuers according to redundant connection and the number of unique domains. Again, both datasets show a considerable overlap: Let's Encrypt (LE) and Google Trust Services (GTS) are the top two issuers w.r.t. redundancy. Jointly, they form the majority of CERT cases mirroring their overall market shares either w.r.t. connections or domains, which we show in detail in Appendix A.1. Hence, for both issuers, we are interested in the involved parties.

We can see differences in the number of occurring unique domains involved in the redundant connections. GTS occurs for fewer domains, but with a high volume of connections, i.e., they are heavy-hitters. In contrast, other issuers do not see such a concentration. Hence, Google alone could significantly reduce the causes for CERT redundancy by changing their certificate issuing policies.

Table 4, showing the top domains involved in CERT cases, underlines this hypothesis. Google ad domains of the top 5 make up 65 % / 63 % of connections with certs issued by GTS. While we see the highest redundancy for an LE-issued domain (accounting for 33 % / 23 % of connections for LE), the distribution is more long-tailed, i.e., the remaining redundancies are spread across many more small websites with potentially different operators. Still, LE is in a position to educate and encourage users, e.g., by nudging users to merge subdomains configured with certbot [7].

**Takeaway.** *Google Trust Services and Let's Encrypt issue the majority of disjunct certificates for redundant connections of cause CERT. Google occurs for few frequently used domains, which are again related to ads. A single party can make a big change. Let's Encrypt, however, occurs for many domains, which are less frequently used. They potentially involve many small website operators who are probably not aware of their certificates disallowing reuse. Quick changes with significant impact are not very likely.*

*5.3.3 CRED: SAN included domains with equal IPs.* The last cause, CRED, represents connections that could be reused w.r.t. HTTP/2 Connection Reuse, but effectively are not. 6 % / 8 % of connections are affected by this scenario; 90 % / 60 % of these even connect to the same domain again.

| Domain | HTTP Archive | | Alexa 100k | | |
|---|---|---|---|---|---|
| | ↑ | Conns. | ↑ | Conns. | Issuer |
| fast.a.klaviyo.com | 1 | 100.31 k | 3 | 1.46 k | LE |
| prev: static.klaviyo.com | | 100.04 k | | 1.46 k | |
| adservice.google.com | 2 | 83.73 k | 2 | 1.56 k | GTS |
| prev: pagead2.googlesyndication.com | | 43.99 k | | 487 | |
| googleads.g.doubleclick.net | 3 | 52.66 k | 6 | 935 | GTS |
| prev: www.googleadservices.com | | 51.05 k | | 692 | |
| pagead2.googlesyndication.com | 4 | 48.43 k | 1 | 1.61 k | GTS |
| prev: adservice.google.com | | 46.82 k | | 797 | |
| images.squarespace-cdn.com | 5 | 45.60 k | | | DCI |
| prev: static1.squarespace.com | | 45.57 k | | | |
| adservice.google.de | 505 | 81 | 4 | 1.37 k | GTS |
| prev: pagead2.googlesyndication.com | | | | 498 | |
| sync.targeting.unrulymedia.com | | | 5 | 1.33 k | DCI |
| prev: sync.1rx.io | | | | 1.33 k | |

**Table 4: Top 5 domains encountered for redundant connections to the same IPs due to absent SAN entries (CERT).**

We suspect the Fetch Standard's credentials flag (cf. Sec. 3) to be responsible for this effect, which has the advantage that only a few browser vendors would have to adapt. To rule out other effects, Figure 2 and Table 1 show results when patching Chromium to ignore this flag (internally named `privacy_mode` [12]).

We can see that the CRED cases vanish completely, but, at first look counter-intuitively, other causes also reduce. We attribute this part to our limitations, but mainly to cases with multiple causes that now disappear. This can also be seen in the absolute difference of all connections which is closer to the CRED difference in contrast to ∼40 k vanished IP cases.

**Takeaway.** *The Fetch Standard's credentials flag is the reason for redundant connections of cause CRED. Disabling it reduces redundancy by 25 %. As only a few parties have to adapt here, a reduction is possible. However, a thorough privacy analysis of Fetch should be conducted, especially given that its added value is already discussed [22].*

## 6 CONCLUSION

In this paper, we study when HTTP/2 Connection Reuse is ineffective and under which circumstances Chromium opens redundant connections, potentially harming performance.

We find that 36 % - 72 % of the 6.24M HTTP Archive and 78 % of the Alexa Top 100k websites open redundant connections. These can be traced back to tracking and advertisements embedding further domains with unsynchronized load-balancing, the Fetch Standard refusing reuse due to (questionable) privacy concerns, and domain sharding with disjunct certificates.

Redundant connections are thus no history and HTTP/3 using the same mechanism will also encounter them. However, we see easy steps for mitigation due to the centricity of the leading causes: Adaption of the Fetch Standard and load-balancing by advertisers need small changes by few parties but can meet a large audience. On the other hand, the merging of certificates for domain sharding includes many parties with relatively small footprints each and will probably take much longer as many operators need to be educated. For future work, we see it as interesting to study the exact performance impact of our findings.

## ACKNOWLEDGMENTS

This work has been funded by the German Research Foundation DFG under Grant No. WE 2935/20-1 (LEGATO). We thank the anonymous reviewers and our shepherd Mattijs Jonker for their valuable comments. We further thank the network operators at RWTH Aachen University, especially Jens Hektor and Bernd Kohler.

| Certificate Issuer | ↑ | HTTP Archive Conns. | Domains | ↑ | Alexa 100k Conns. | Domains |
|---|---|---|---|---|---|---|
| Google Trust Services | 1 | 28.14 M | 259.50 k | 1 | 844.42 k | 11.45 k |
| DigiCert Inc | 2 | 12.12 M | 326.24 k | 2 | 291.77 k | 21.98 k |
| Cloudflare, Inc. | 3 | 6.44 M | 698.13 k | 3 | 136.58 k | 31.57 k |
| Amazon | 4 | 4.74 M | 256.03 k | 4 | 135.71 k | 18.20 k |
| Let's Encrypt | 5 | 4.38 M | 1.89 M | 5 | 70.64 k | 28.33 k |
| Sectigo Limited | 6 | 3.59 M | 434.18 k | 6 | 64.63 k | 10.62 k |
| GlobalSign nv-sa | 7 | 1.52 M | 115.13 k | 7 | 47.08 k | 6.01 k |
| GoDaddy.com, Inc. | 8 | 814.81 k | 134.92 k | 8 | 20.30 k | 4.07 k |
| Yandex LLC | 9 | 468.86 k | 178 | 9 | 9.97 k | 49 |
| COMODO CA Limited | 10 | 250.47 k | 9.92 k | 11 | 4.22 k | 352 |
| Microsoft Corporation | 11 | 239.44 k | 3.03 k | 10 | 8.64 k | 230 |

**Table 5: Top 10 certificate issuers by Issuer Org. for all connections and their original / SNI domain.**

| AS | ↑ | HTTP Archive Conns. | Domains | ↑ | Alexa 100k Conns. | Domains |
|---|---|---|---|---|---|---|
| GOOGLE | 1 | 8.04 M | 143.49 k | 1 | 315.57 k | 7.14 k |
| AMAZON-02 | 2 | 1.74 M | 46.71 k | 2 | 50.67 k | 7.51 k |
| FACEBOOK | 3 | 1.63 M | 360 | 3 | 32.04 k | 81 |
| AUTOMATTIC | 4 | 402.71 k | 3.36 k | 10 | 3.38 k | 58 |
| CLOUDFLARENET | 5 | 307.64 k | 14.22 k | 4 | 9.18 k | 3.40 k |
| FASTLY | 6 | 228.12 k | 1.62 k | 9 | 3.50 k | 1.00 k |
| AMAZON-AES | 7 | 220.43 k | 10.71 k | 5 | 7.22 k | 886 |
| EDGECAST | 8 | 182.22 k | 997 | 6 | 4.33 k | 201 |
| AKAMAI-ASN1 | 9 | 144.52 k | 2.30 k | 8 | 3.53 k | 773 |
| AKAMAI-AS | 10 | 130.77 k | 2.75 k | 7 | 3.61 k | 736 |

**Table 6: Top 10 ASNs for connections of cause IP.**

## A FURTHER RESULTS

In the following, we describe further details about our datasets complementing our results. Here, we look at the total share of issuers for all connections, which ASNs have been observed being affected, the overlap between both datasets and how load-balancing influences the DNS resolving process in greater depth.

### A.1 Certificate Issuer Share

We showed that in the ecosystem of certificate issuers, mainly Google Trust Services and Let's Encrypt are involved for redundant connections of cause CERT. These connections have to be opened due to non-overlapping certificate subject names. In the following, we present the shares of the top 10 certificate issuers over all opened connections in Table 5 to allow these results to be set into perspective. We see that Google Trust Services is the Issuer which occurs for the most connections, while Let's Encrypt is less often seen over all connections, although it is also in the top-2 for the redundant connections of class CERT. Domain-wise, we can see that Let's Encrypt leads, followed by Cloudflare in the HTTP Archive and vice versa for our measurements. All in all, we can see that both leaders, Let's Encrypt w.r.t. domains and Google w.r.t. connections, reflect in our measurements w.r.t. to their market share of overall connections and domains.

### A.2 ASs Affected by Cause IP

Similarly, we showed which domains have been involved in cause IP (cf. 5.3.1), but not which content providers were responsible for providing these resources. Hence, Table 6 shows the top 10 ASs which were involved in redundant connections of type IP. Unsurprisingly, the Google and the Facebook AS show up, which we identified as being mainly responsible for redundant connections of type IP. However, these parties operate their own CDNs, which mostly also provide their content. Contrasting, Amazon shows up in the AS list, but not as an involved party w.r.t. the domains, hinting at its cloud instances or Cloudfront CDN being responsible. Indeed, Hotjar.com (occurs as place 12 for the HTTP Archive measurements / as place 16 for our measurements of the involved domains for the IP case, cf. Table 12) uses the Cloudfront CDN, which distributes static.hotjar.com and script.hotjar.com to different IPs. Amazon adapting its DNS load-balancing could help here to support connection reuse.

| Cause | IP | Cert | HAR Overlap Endless Sites | Conns. | Alexa Overlap Endless Sites | Conns. |
|---|---|---|---|---|---|---|
| CERT | = | ≠ | 4.03 k | 6.63 k | 5.06 k | 8.86 k |
| IP | ≠ | = | 23.85 k | 110.64 k | 27.06 k | 175.05 k |
| CRED | = | = | 16.84 k | 29.61 k | 24.00 k | 51.72 k |
| Redund. | | | 25.34 k | 135.33 k | 28.63 k | 218.97 k |
| Total | | | 29.53 k | 451.71 k | 29.53 k | 608.76 k |

**Table 7: Occurring causes for the overlap / intersection of the HTTP Archive and our measurements.**

The other CDNs and providers occur in much lower quantities. AUTOMATTIC can be related to the Wordpress tools from wp.com, hence its reduced domain count in contrast to Cloudflare. Nevertheless, both account for around and more than 20 times fewer connections than Google, respectively.

### A.3 Overlap of Results Between Both Datasets

Throughout our paper, we compare the results between the HTTP Archive and our Alexa list, which diverge in visited domains. Hence, we here present the overlap of both datasets, i.e., we intersect the datasets w.r.t. the visited URLs to show how both measurements map to each other. The general results are shown in Table 7 for our overlapping 29.53 k sites. As can be seen, all numbers for the Alexa Overlap dataset are larger than for the HAR dataset as we had to filter 490.32 k requests of the total 2.71 M HTTP/2 requests. Our own dataset consists of 2.98 M HTTP/2 requests, of which none had to be filtered. Table 8 shows the top 5 origins for the IP cause for the overlap. It matches the top origins in Table 2 surprisingly well. However, we can still see a difference in the top origins between our and the HTTP Archive measurements which cannot be explained by the filtered requests. Instead, our geolocation seems to affect Google to redirect us to its German domain. Nevertheless, the remaining origins differ only slightly in their connection counts. Table 9 shows the top certificate issuers between both measurements.

| | HTTP Archive | | Alexa | |
|---|---|---|---|---|
| Origin | ↑ | Conns. | ↑ | Conns. |
| www.google-analytics.com | 1 | 15.34 k | 1 | 20.76 k |
| prev: www.googletagmanager.com | | 14.74 k | | 14.59 k |
| www.facebook.com | 2 | 9.44 k | 4 | 9.36 k |
| prev: connect.facebook.net | | 9.44 k | | 9.34 k |
| pagead2.googlesyndication.com | 3 | 6.52 k | 5 | 7.26 k |
| prev: www.googletagservices.com | | 4.12 k | | 5.57 k |
| googleads.g.doubleclick.net | 4 | 5.91 k | 6 | 7.24 k |
| prev: pagead2.googlesyndication.com | | 4.20 k | | 4.94 k |
| tpc.googlesyndication.com | 5 | 5.15 k | 9 | 6.05 k |
| prev: pagead2.googlesyndication.com | | 3.49 k | | 4.90 k |
| www.google.de | | | 2 | 10.16 k |
| prev: www.gstatic.com | | | | 6.97 k |
| apis.google.com | 79 | 106 | 3 | 9.71 k |
| prev: www.gstatic.com | | | | 9.64 k |

**Table 8: Top 5 origins, their redundant connections, rank (↑) and reusable previous connections for cause IP for the overlap / intersection of the HTTP Archive and our measurements.**

| | HTTP Archive | | | Alexa | | |
|---|---|---|---|---|---|---|
| Certificate Issuer | ↑ | Conns. | Domains | ↑ | Conns. | Domains |
| Google Trust Services | 1 | 2.69 k | 130 | 1 | 3.63 k | 73 |
| Let's Encrypt | 2 | 2.39 k | 733 | 2 | 2.42 k | 772 |
| DigiCert Inc | 3 | 655 | 189 | 3 | 1.34 k | 213 |
| Sectigo Limited | 4 | 202 | 110 | 5 | 308 | 118 |
| GlobalSign nv-sa | 5 | 189 | 100 | 4 | 347 | 118 |

**Table 9: Top 5 certificate issuers w.r.t. redundant connections of cause CERT and unique domains for the overlap / intersection of the HTTP Archive and our measurements.**

Connection-wise, the results are very close and the number of domains also differs by less than a factor of 2. Still, the numbers do not match perfectly, which we attribute to the vantage point difference. Similarly, Table 10 shows that the top domains for cause CERT overlap, but are not exactly the same and again our geolocation seems to influence redirection at Google.

### A.4 Load-Balancing Influence on Cause IP

Moreover, we saw that certain domains point to different IPs although the underlying content was available at the respectively other IP. We attribute this effect to load-balancing, which can also be a root-cause for differences between our measurements and the HTTP Archive measurements. Hence, we check the temporal and spatial dependency of DNS resolution for our 20 most-occurring domains (which we present in Table 12). We resolve these domains over the course of several days via 14 different DNS resolvers, which we list in Table 11. We selected these resolvers via public DNS lists at https://dnschecker.org/ and https://public-dns.info/ and made sure that they had reverse DNS entries. Moreover, we checked that ECS [6] is not supported. We then queried the domains every 6 minutes and checked whether the resulting IPs overlap, i.e., whether connection reuse is aided. In Figure 3, we show the temporal pattern of the overlaps by counting the resolvers for which the two domains shown aside resolve to overlapping IPs. We filtered time slots where not all DNS resolvers answered correctly to avoid noise due to missing data. Darker areas denote more resolvers for which the DNS answers overlapped. We can see that certain domains never overlap, while other domains fluctuate. E.g., www.google-analytics.com and www.googletagmanager.com did not overlap, while fonts.gstatic.com and gstatic.com overlap sometimes. I.e., time and also vantage point influencing the load-balancing influence whether domains resolve to the same IP and connection reuse is effective or not.

| | HTTP Archive | | Alexa | | |
|---|---|---|---|---|---|
| Domain | ↑ | Conns. | ↑ | Conns. | Issuer |
| adservice.google.com | 1 | 864 | 2 | 661 | GTS |
| prev: pagead2.googlesyndication.com | | 397 | | 253 | |
| fast.a.klaviyo.com | 2 | 609 | 3 | 641 | LE |
| prev: static.klaviyo.com | | 608 | | 641 | |
| pagead2.googlesyndication.com | 3 | 481 | 1 | 695 | GTS |
| prev: adservice.google.com | | 470 | | 351 | |
| googleads.g.doubleclick.net | 4 | 412 | 7 | 307 | GTS |
| prev: www.googleadservices.com | | 393 | | 246 | |
| alb.reddit.com | 5 | 161 | 11 | 185 | DCI |
| prev: www.redditstatic.com | | 161 | | 185 | |
| adservice.google.de | | | 4 | 620 | GTS |
| prev: pagead2.googlesyndication.com | | | | 258 | |
| sync.targeting.unrulymedia.com | | | 5 | 533 | DCI |
| prev: sync.1rx.io | | | | 533 | |

**Table 10: Top 5 domains encountered for redundant connections to the same IPs due to absent SAN entries (CERT) for the overlap / intersection of the HTTP Archive and our measurements.**

| IP | Country | Operator |
|---|---|---|
| internal | Germany | RWTH Aachen University |
| 168.126.63.1 | South Korea | KT Corporation |
| 172.104.237.57 | Germany | FreeDNS |
| 172.104.49.100 | Singapore | FreeDNS |
| 177.47.128.2 | Brazil | Ver Tv Comunicações S/A |
| 178.237.152.146 | Spain | MAXEN TECHNOLOGIES, S.L. |
| 195.208.5.1 | Russia | MSK-IX |
| 203.50.2.71 | Australia | Telstra Corporation Limited |
| 210.87.250.59 | Hong Kong | HKT Limited |
| 212.89.130.180 | Germany | Infoserve GmbH |
| 221.119.13.154 | Japan | Marss Japan Co., Ltd |
| 8.0.26.0 | United Kingdom | Level 3 Communications, Inc. |
| 8.0.6.0 | USA | Level 3 Communications, Inc. |
| 80.67.169.12 | France | French Data Network (FDN) |

**Table 11: DNS resolvers used to analyze DNS-based load-balancing.**

| Origin | HTTP Archive | | Alexa 100k | |
|---|---|---|---|---|
| | ↑ | Conns. | ↑ | Conns. |
| www.google-analytics.com | 1 | 2.25 M | 1 | 52.31 k |
| prev: www.googletagmanager.com | | 2.12 M | | 36.93 k |
| www.facebook.com | 2 | 1.52 M | 4 | 25.05 k |
| prev: connect.facebook.net | | 1.51 M | | 25.02 k |
| googleads.g.doubleclick.net | 3 | 615.45 k | 6 | 17.03 k |
| prev: pagead2.googlesyndication.com | | 398.41 k | | 10.19 k |
| pagead2.googlesyndication.com | 4 | 606.59 k | 7 | 16.99 k |
| prev: googleads.g.doubleclick.net | | 418.69 k | | |
| prev: www.googletagservices.com | | | | 12.60 k |
| tpc.googlesyndication.com | 5 | 465.56 k | 9 | 13.38 k |
| prev: pagead2.googlesyndication.com | | 362.79 k | | 9.93 k |
| www.gstatic.com | 6 | 439.01 k | 185 | 117 |
| prev: fonts.gstatic.com | | 330.07 k | | 97 |
| www.googletagservices.com | 7 | 416.76 k | 13 | 9.76 k |
| prev: pagead2.googlesyndication.com | | 370.60 k | | 8.21 k |
| partner.googleadservices.com | 8 | 379.79 k | 15 | 7.05 k |
| prev: pagead2.googlesyndication.com | | 376.37 k | | 7.04 k |
| www.google.com | 9 | 379.12 k | 442 | 34 |
| prev: www.youtube.com | | 126.03 k | | |
| stats.g.doubleclick.net | 10 | 342.53 k | 8 | 16.02 k |
| prev: googleads.g.doubleclick.net | | 203.38 k | | 10.96 k |
| fonts.gstatic.com | 11 | 231.71 k | 64 | 634 |
| prev: www.gstatic.com | | 65.16 k | | 582 |
| script.hotjar.com | 12 | 226.57 k | 16 | 6.98 k |
| prev: static.hotjar.com | | 225.67 k | | 6.98 k |
| vars.hotjar.com | 13 | 216.27 k | 18 | 6.76 k |
| prev: static.hotjar.com | | 213.04 k | | 6.76 k |
| in.hotjar.com | 14 | 202.01 k | 20 | 3.98 k |
| prev: static.hotjar.com | | 196.92 k | | |
| prev: script.hotjar.com | | | | 3.98 k |
| fonts.googleapis.com | 15 | 186.71 k | 482 | 29 |
| prev: ajax.googleapis.com | | 152.37 k | | 19 |
| stats.wp.com | 16 | 157.30 k | 36 | 1.28 k |
| prev: c0.wp.com | | 73.49 k | | 908 |
| securepubads.g.doubleclick.net | 17 | 143.31 k | 17 | 6.78 k |
| prev: www.googletagservices.com | | 112.36 k | | 5.76 k |
| ajax.googleapis.com | 18 | 142.01 k | 12 | 11.00 k |
| prev: fonts.googleapis.com | | 136.17 k | | 11.00 k |
| maps.googleapis.com | 19 | 118.04 k | 26 | 2.59 k |
| prev: fonts.googleapis.com | | 109.55 k | | 2.59 k |
| www.googleadservices.com | 20 | 106.64 k | 21 | 3.79 k |
| prev: stats.g.doubleclick.net | | 78.97 k | | 2.31 k |
| www.google.de | 12 255 | 6 | 2 | 27.74 k |
| prev: www.gstatic.com | | | | 18.61 k |
| apis.google.com | 82 | 14.25 k | 3 | 26.10 k |
| prev: www.gstatic.com | | | | 25.91 k |
| ogs.google.com | 1 375 | 14.25 k | 5 | 26.10 k |
| prev: www.gstatic.com | | | | 19.03 k |
| adservice.google.com | 22 | 95.16 k | 10 | 12.88 k |
| prev: www.gstatic.com | | | | 8.93 k |
| adservice.google.de | 4 944 | 25 | 11 | 12.68 k |
| prev: www.gstatic.com | | | | 8.44 k |
| cm.g.doubleclick.net | 28 | 69.85 k | 14 | 7.71 k |
| prev: googleads.g.doubleclick.net | | | | 5.67 k |
| i.ytimg.com | 115 | 9.32 k | 19 | 4.25 k |
| prev: www.gstatic.com | | | | 2.66 k |

**Table 12: Top 20 domains for the IP case.**

# of resolvers with overlap

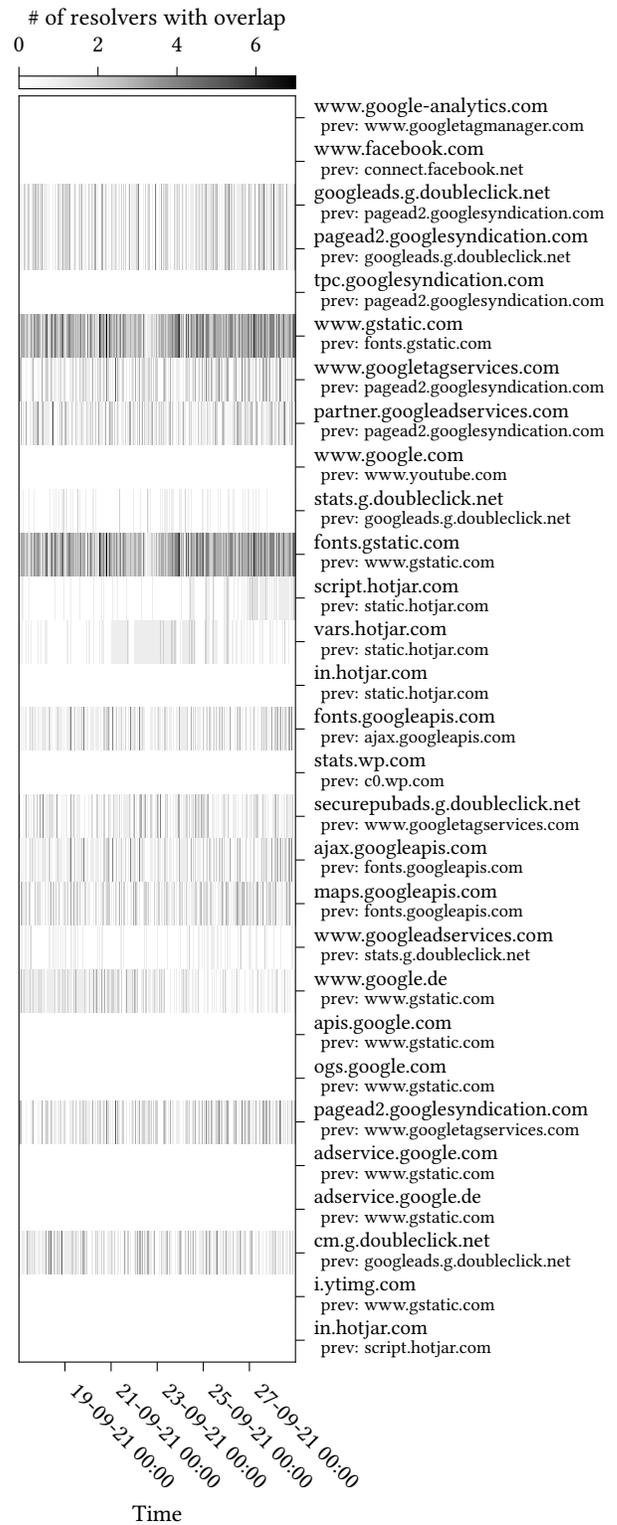

**Figure 3: Number of DNS vantage points where domains overlapped.**